\documentclass[twocolumn]{autart}   

\usepackage{amssymb}
\usepackage{amsmath}
\usepackage{amsfonts}
\usepackage{indentfirst}
\usepackage{lipsum}[1]
\usepackage[longnamesfirst]{natbib}                     
\usepackage[latin1] {inputenc}
\usepackage{enumerate}
\usepackage{mathrsfs}
\usepackage[breaklinks=true]{hyperref}
\usepackage{breakcites}
\usepackage{tabulary}
\usepackage{cite}
\usepackage{wrapfig}
\usepackage{psfrag}
\usepackage{graphicx}          
\usepackage{color}      
\usepackage{epsfig} 
\usepackage{times} 

\def\qedp{\hspace*{\fill}~{\tiny $\blacksquare$}}
\def\qed{\relax\ifmmode\hskip2em \Box\else\unskip\nobreak\hskip1em $\Box$\fi}

\newtheorem{theorem}{Theorem}
\newtheorem{itlemma}{Lemma}
\newtheorem{itdefinition}{Definition}
\newtheorem{itproposition}{Proposition}
\newtheorem{itresult}{Result}
\newtheorem{itremark}{Remark}
\newtheorem{itassumption}{Assumption}
\newtheorem{itcorollary}{Corollary}
\newtheorem{itexample}{Example}
\newenvironment{definition}{\begin{itdefinition}\rm}{\end{itdefinition}}

\newenvironment{corollary}{\begin{itcorollary}\rm}{\end{itcorollary}}

\begin{document}

\begin{frontmatter}

\title{Low-complexity Learning of Linear Quadratic Regulators \\
from Noisy Data}

\vspace{-0.5cm}
\author[RUG]{Claudio De Persis}  
\author[Florence]{Pietro Tesi} 

\address[RUG]{
ENTEG, Faculty of Science and Engineering, University of
Groningen, 9747 AG Groningen, The Netherlands}      
\address[Florence]{
Department of Information Engineering (DINFO), University of Florence,
50139 Florence, Italy}              
	
\begin{abstract}
This paper considers the Linear Quadratic Regulator 
problem for linear systems with unknown dynamics, a central problem in data-driven 
control and reinforcement learning. We propose a method that
uses data to directly return a controller without estimating a model of the system. 
Sufficient conditions are given under which this method returns
a stabilizing controller with guaranteed relative error when the data
used to design the controller are affected by noise.
This method has low complexity as it only requires a finite 
number of samples of the system response to a sufficiently 
exciting input, and can be efficiently implemented as a semi-definite program.
Further, the method does not require assumptions on the noise statistics, and the 
relative error nicely scales with the noise magnitude. % or signal-to-noise ratio.
\end{abstract}
	
\end{frontmatter}	

\section{Introduction}

Control theory is witnessing an increasing renewed
interest towards \emph{data-driven} (\emph{data-based}) control. 
This terminology refers to all those cases where
the dynamics of the system are unknown and the control law must 
be designed using data alone. 
This can be done either by
identifying a model of the system from data and then use the model for control design,
or by directly designing the control law bypassing the
system identification (ID) step. Methods in the first 
category are usually called \emph{indirect} (sequential system ID
and control design), while methods in the second 
category are usually called \emph{direct} or \emph{model-free}. \\[0.2cm]
The interest for data-driven control has several motivations.
As systems become more complex, first-principle
models may be difficult to obtain or may be too complex for control
design. Fully automated (end-to-end) procedures may also facilitate
the online tuning or re-design of controllers, which is needed in all 
those applications where the system to be controlled or the environment
are subject to changes that are difficult to predict. 
Dozens of publications on data-driven control have appeared in
the last few years. We mention works on 
predictive control \citep{Alpago2020,Coulson2018,Salvador2018}, optimal control \citep{Goncalves2019,Baggio2019,recht_annual,deptCDC}, 
robust and nonlinear control \citep{dept2020,Berberich2019,Zeilinger2019,Dai2018,D2IBC}.
This list is by no means exhaustive. We refer the interested reader to 
\citep{Hou2013} for a survey on earlier contributions. 

\emph{The Linear Quadratic Regulator problem} 

This paper considers the \emph{infinite horizon} Linear Quadratic Regulator (LQR) problem
for linear time-invariant systems, which is
one of the problems more studied in the control literature. 
Besides its practical relevance, this problem is a prime example 
of the challenges encountered in data-driven control. 
Specifically, we consider the problem of computing the 
solution to the LQR problem from a finite 
set of ({noisy}) data collected from the system. \\[0.2cm]
Early data-driven methods for LQR can be traced back to 
the theory of adaptive control systems, and include the popular  
\emph{self-tuning regulators} \citep{Astrom1989} and 
\emph{policy iteration} schemes \citep{Bradtke1994}. While the specific 
techniques are different, the common idea is to study the convergence of an adaptive control law
to the optimal one as time goes to infinity. Starting from \citep{Fiechter1997},
a tremendous effort has been made for establishing 
\emph{non-asymptotic} properties of data-driven methods.
This term refers to all those methods that aim at providing 
closed-loop stability and performance guarantees using 
only a \emph{finite} number of data points.
The interest towards \emph{non-asymptotic} properties is both theoretical  
and practical. Non-asymptotic properties
help to derive performance guarantees of iterative (online) methods 
\citep{Fazel20182385}, and are at the basis of non-iterative (offline) methods
\citep{recht_annual,deptCDC}.\footnote{Here, by \emph{iterative} we refer to
all those methods where the control law is modified online.} \\[0.2cm]
It turns out that non-asymptotic properties are very difficult to derive
if one departs from the assumption that data are noise-free.
Most of the works dealing with noisy data are of indirect type and come from the area of
\emph{reinforcement learning} (RL). The common approach is to learn a model of
the system along with non-asymptotic \emph{probabilistic} bounds 
on the estimation error \citep{CampiMC}, and then design or update the control
law depending on the specific method adopted (non-iterative or iterative).
Among iterative methods we mention \citep{Cohen2019,Abbasi2018},
where the latter is one of the few model-free methods that appeared in
the literature. Among non-iterative methods we mention
\citep{Mania2019,Dean2018journ}. 

\emph{Our contribution}

Our contribution is a new approach 
to design LQ controllers from noisy data with
guaranteed performance. The method
builds on the framework introduced in \citep{dept2020}
and has the following features:
\begin{enumerate}
\item[(i)] \emph{Low complexity}. 
The proposed method requires 
a finite (pre-computable) number of data points obtained from a 
single or multiple system's trajectories, and it can be 
implemented as a convex program.
\item[(ii)] \emph{Stability and performance guarantees}. 
As long as the noise 
satisfies suitable inequalities
our method returns a stabilizing controller with quantitative
\emph{relative error} (gap between the computed solution and the 
unknown optimal controller)
and the error nicely scales with the noise magnitude.
\item[(iii)] \emph{No assumptions on statistical properties of noise}. 
We do not make assumptions
regarding the noise statistics such as the noise being a martingale or white.  
\end{enumerate}

As in \citep{recht_annual,Mania2019,Dean2018journ}, we focus on
non-iterative methods which do not require an initial stabilizing
controller, as instead typically assumed in iterative methods 
 \citep{Cohen2019,Abbasi2018}. The main difference with
 respect to \citep{recht_annual,Mania2019,Dean2018journ}
 is that our method is direct and assumes no noise model. \\[0.2cm]
The advantage of not relying on noise statistics is twofold.
Although the solution to LQR can be interpreted as the one 
minimizing the variance of the system's output in response to 
white noise, experimental data need not comply with 
such setting, and show correlation and dependence
(dependence breaks the \emph{i.i.d.} assumption
used in \citep{Mania2019,Dean2018journ}). 
Our method is free from this issue,
while it also allows for simple noise-reduction strategies
for random noise.
Not relying on noise statistics also enables us to directly 
extend the analysis to the stabilization of nonlinear systems
around an equilibrium since, around an equilibrium point, a nonlinear system 
can be expressed via its first-order approximation plus a remainder
which acts as a noise source.
We will elaborate on these point in the paper.\\[0.2cm]
Our method is direct (model-free). There is currently a great debate 
regarding the effectiveness of direct methods versus indirect ones \citep{recht_comp}. 
Here, we will not enter this debate partly because existing methods for LQR give
{probabilistic} results while our method is \emph{non-probabilistic},
and since the question of what is the ``best" approach to take
remains a question on the \emph{priors}. 
A strength of our method (of direct methods in general)
is a parsimonious use of such priors, which allows us to 
cope with situations where the noise has no convenient statistics.
In such situations indirect methods (at least those proposed for LQR) 
are instead much more difficult to pursue 
since the ID step is strongly reliant on such statistics \citep{Mania2019,Dean2018journ}. 
On a similar vein,
direct methods can be directly applied in settings such as with 
nonlinear or time-varying dynamics where system ID is typically
more involved 
\citep{Zeilinger2019,Dai2018,dept2020,Guo2020}.
Finally, by skipping the ID step, 
direct methods are often much
more handy than indirect methods. 
For instance, regarding the LQR problem, 
our method does not require any \emph{bootstrap} 
method or \emph{reset} of the system's state, as needed in
\citep{Mania2019,Dean2018journ}. 

\emph{Outline of the paper}

Our method rests on a fundamental result by Willems and 
co-authors \citep{willems2005note} recalled in Section \ref{Sec2}. 
Roughly, this result states that a (noise-free) system trajectory generated by 
a \emph{persistently exciting} input is a data-based non-parametric system model.
We exploit this result to develop our model-free method.
In Section \ref{Sec3}, we formulate the LQR as an $\mathcal H_2$
problem \citep{scherer-LMI-book} and derive a data-based solution 
based on convex programming 
for the ideal case of noise-free data (Theorem \ref{thm:main1}). 
The main results are 
given through  
Sections \ref{Sec4} and \ref{Sec5}. The first one (Theorem \ref{thm:main2})
provides stability properties and error bounds of the baseline solution in
case of noisy data. 
Two variants to the baseline solution are discussed 
in Theorems \ref{thm:main3} and \ref{thm:main4}. 
These variants guarantee more tolerance to noise at the cost of
possibly reduced performance bounds.
This matches what has been observed in indirect methods \citep{Mania2019}
for noise-robust solutions.
Practical aspects and extensions are discussed in Section \ref{Sec6},
including nonlinear systems and de-noising strategies.
Section \ref{Sec7} provides numerical simulations while
Section \ref{Sec8} gives concluding remarks.

\section{Notation and auxiliary facts} \label{Sec2}

Given a signal $z: \mathbb{N}\to \mathbb{R}^\sigma$ and two 
integers $k$ and $r$ where $r \geq k$ we define $z_{[k,r]}:=\{ z(k),z(k+1),\ldots,z(r) \}$.
Given a signal $z$ and 
a positive integer $T$,
we define
\begin{eqnarray*}
&& Z_i = Z_{i,T} := \left[ \begin{array}{cccc} z(i) & z(i+1) & \cdots & z(T+i-1) \end{array} \right] 
\end{eqnarray*}
As we will always consider data coming from experiments of 
length $T$, we write $Z_i$ instead of $Z_{i,T}$ for brevity. \\[0.2cm]
Consider a linear time-invariant system
\begin{eqnarray} \label{eq:system0}
x(k+1) = A x(k) + B u(k) \quad k \in \mathbb N
\end{eqnarray}
where $x \in \mathbb R^n$ is the state and $u \in \mathbb R^m$ is the control input,
and suppose that we have access to $T$-long data sequences $u_{[0,T-1]}$ and $x_{[0,T-1]}$
of system \eqref{eq:system0}.
Throughout the paper, the condition 
\begin{eqnarray} \label{rank.condition}
{\rm rank} \, W_0 = n + m
\end{eqnarray}
where
\begin{eqnarray} \label{}
W_0 :=
\left[
\begin{array}{c}
U_{0} \\
X_{0}
\end{array}
\right]
\end{eqnarray}
plays an important role. 
Condition \eqref{rank.condition} guarantees that
any $T$-long input-state trajectory of the system can be expressed as a linear combination 
of the columns of $W_{0}$, meaning that $W_0$ encodes all the information
regarding the dynamics of the system.
A fundamental property established in \citep{willems2005note}
is that one can guarantee \eqref{rank.condition} 
when the input is sufficient \emph{exciting}. This property has very much in common with the notion
of \emph{active exploration} \citep{Fiechter1997} used in many reinforcement 
learning methods.

\begin{definition}
{\rm \citep{willems2005note}}
A signal $z_{[0, T-1]}\,\in \mathbb R^\sigma$ 
is said \emph{persistently exciting} of order $s \in \mathbb N_1$ if the matrix 
\begin{eqnarray*}
\mathscr Z_0 = \mathscr Z_{0,T} :=
\begin{bmatrix}
z(0) & z(1) & &\cdots && z(T-s)\\
z(1) & z(2) & &\cdots && z(T-s+1)\\
\vdots &\vdots &&\ddots && \vdots \\
z(s-1) & z(s) & & \cdots && z(T-1)
\end{bmatrix}
\end{eqnarray*}
has full rank $\sigma s$. \qedp
\end{definition}

\begin{lem}\label{lem:willems}
{\rm \citep[Corollary 2]{willems2005note}}
Suppose that system \eqref{eq:system0} is controllable.
If $u_{[0,T-1]}$ 
is persistently exciting of order $n+1$, then condition \eqref{rank.condition} holds. \qedp
\end{lem} 

\section{Problem definition and data-driven formulation} \label{Sec3}

In this section, we introduce the problem of interest and our
baseline direct (model-free) data-driven method, which rests
on condition \eqref{rank.condition} above.

\subsection{The Linear Quadratic Regulator problem}

Consider a linear time-invariant system 
{\setlength\arraycolsep{2pt} 
\begin{eqnarray} \label{eq:system}
\left\{
\def\arraystretch{1.3}
\begin{array}{rl}
x(k+1) &= Ax(k)+Bu(k)+d(k) \\[0.2cm]
z(k) &= \left[
\begin{array}{cc}
W_x^{1/2} & 0 \\ 0 & W_u^{1/2}
\end{array}
\right] \left[
\begin{array}{c}
x(k) \\ u(k)
\end{array}
\right] 
\end{array}
\right.
\end{eqnarray}}%
where $x \in \mathbb R^n$ is the state, $u \in \mathbb R^m$ is the control input,
and where $d$ is a disturbance term;
$z$ is a performance signal of interest; $(A,B)$ is controllable;
$W_x \succeq 0$ and $W_u \succ 0$ are weighting matrices with $(W_x,A)$ observable.
In the sequel, to simplify the notation we set  $W_x = W_u = I$, although all 
the results easily extend to the general case. \\[0.2cm]
We consider the problem of designing a state-feedback controller $K$ that renders
$A+BK$ Hurwitz and minimizes 
the $\mathcal H_2$-norm of the transfer function $\mathscr T(K): d \rightarrow z$
of the closed-loop system
\begin{eqnarray} \label{eq:closed}
\left[
\begin{array}{c}
x(k+1) \\ z(k)
\end{array}
\right]  = 
\left[
\begin{array}{c|c}
A+BK & I \\ \hline 
\left[ \begin{array}{c} I \\ K \end{array} \right]  & 0
\end{array}
\right] \left[
\begin{array}{c}
x(k) \\ d(k)
\end{array}
\right] 
\end{eqnarray}
where \citep[Section~4.4]{Chen1995}
\begin{eqnarray*}
\| \mathscr T(K) \|_2 := \left[ \frac{1}{2 \pi} \int_{0}^{2 \pi} {\rm trace} 
\left( h \left(e^{j \theta}\right)^\top h \left(e^{j \theta}\right) \right) d\theta \right]^{\frac{1}{2}}
\end{eqnarray*}
In particular \citep[Section~4.4]{Chen1995}, when
$A+BK$ is Hurwitz,
\begin{eqnarray} \label{eq:trace_cost}
\| \mathscr T(K) \|^2_2 = {\rm trace}\left( P \right) + {\rm trace}\left( K P K^\top \right) 
\end{eqnarray}
where $P$ is the controllability Gramian of the closed-loop system \eqref{eq:closed}, 
which is the unique solution to
\begin{eqnarray}
(A+BK) P (A+BK)^\top - P + I = 0
\end{eqnarray}
This corresponds in the time domain to the $2$-norm 
of the output $z$ when impulses are applied to the input channels, and can
be interpreted as the mean-square deviation of $z$ when 
$d$ is a white process with unit covariance, which is the classic stochastic
LQR formulation. 
Here, we view the LQR problem as a $\mathcal H_2$-norm minimization 
problem as our method is based on the minimization of \eqref{eq:trace_cost}. \\[0.2cm]
It is known \citep[Section 6.4]{Chen1995} that the state-feedback 
controller which minimizes the $\mathcal H_2$-norm of $\mathscr T(K)$
is unique and 
can be computed as
\begin{eqnarray}
K_{opt} = - (I+ B^\top X B)^{-1} B^\top X A
\end{eqnarray} 
where $X$ is the unique  positive definite solution to the 
classic discrete-time algebraic Riccati (DARE) equation
{\setlength\arraycolsep{2pt}
\begin{eqnarray*} \label{eq:H2_Riccati}
&& A^\top X A - X \\ 
&& \quad - (A^\top X B)(I+B^\top X B)^{-1} (B^\top X A )+I = 0
\nonumber 
\end{eqnarray*}}%
We are interested in computing $K_{opt}$ when a model
of the system is not available, and we only have access to a $T$-long
stream of (nosy) data $u_{[0,T-1]}$ and $x_{[0,T-1]}$
collected during some experiment on system \eqref{eq:system}. By \emph{noisy}
we mean that the data collected from \eqref{eq:system} might have been
generated with \emph{nonzero} disturbance $d$. In particular, we aim at establishing 
properties of the data-driven solution with respect to the one
that we can compute under exact model knowledge. 

\subsection{A data-driven SDP formulation}

The problem of finding $K_{opt}$ 
can be equivalently formulated as a \emph{semi-definite program} (SDP):\footnote{With
some abuse of terminology, we refer to \eqref{lqr-form_model} and subsequent
derivations as an SDP, with the understanding that they can be written as SDP
using standard manipulations.}
\begin{eqnarray} \label{lqr-form_model}
\begin{array}{l}
\min_{(\gamma,K,P,L)}  \,\, \gamma \\ 
\textrm{subject to} \smallskip \smallskip \\
\left\{
\def\arraystretch{1.3} 
\begin{array}{l}
(A+BK) P (A+BK)^\top  - P + I \preceq 0 \\ 
P \succeq I \\
L - K P K^\top \succeq 0 \\
{\rm trace}\left( P \right) + {\rm trace} \left( L \right) \leq \gamma
\end{array} \right.
\end{array}
\end{eqnarray}
This formulation is the natural discrete-time counterpart of the formulation
proposed in \citep{feron1992numerical} for continuous-time systems. 
We will not discuss the properties associated to \eqref{lqr-form_model}.
Rather, we will discuss the properties associated to an \emph{equivalent 
data-based} version of \eqref{lqr-form_model}. \\[0.2cm]
Consider system \eqref{eq:system} along with data sequences
$d_{[0,T-1]}$, $u_{[0,T-1]}$ and $x_{[0,T]}$ resulting from
an experiment of length $T$. Define corresponding matrices $D_0$, $U_0$,
$X_0$ and $X_1$, which satisfy the relation
\begin{eqnarray} \label{eq:recursive}
X_1 = A X_0 + B U_0 + D_0
\end{eqnarray}
It turns out that the controller $K_{opt}$ can
be parametrized directly in terms of the data matrices
$D_0$, $U_0$, $X_0$ and $X_1$.
Specifically, under condition \eqref{rank.condition} the controller  
$K_{opt}$ can
be expressed as 
\begin{eqnarray} \label{eq:Ko_data}
K_{opt}=U_0Q_oP_o^{-1}
\end{eqnarray}
where the tuple $(\gamma_o,Q_o,P_o,L_o)$ is any optimal solution to the SDP:
\begin{eqnarray} \label{lqr-form_ideal}
\begin{array}{l}
\min_{(\gamma,Q,P,L)} {\gamma}  \\
\textrm{subject to} \smallskip \smallskip \\
\left\{
\def\arraystretch{1.3} 
\begin{array}{l}
(X_{1} -D_0) Q P^{-1} Q^\top (X_{1}-D_0)^\top  - P + I \preceq 0 \\ 
P \succeq I \\
L - U_{0}Q P^{-1} Q^\top U_{0}^\top \succeq 0 \\
X_{0} Q = P \\
 {\rm trace}\left( P \right) + {\rm trace} \left( L \right) \leq \gamma
\end{array} \right.
\end{array}
\end{eqnarray}
which only depends on data. \\[0.2cm]
The idea behind this formulation is that, under condition \eqref{rank.condition}, 
any feedback interconnection $A+BK$ can be rewritten in a form which does not involve 
the matrices $A$ and $B$. In fact, under condition \eqref{rank.condition}, 
for any $K$ there exists a matrix $G$ 
that solves the system of equations
\begin{eqnarray} \label{eq:constG}
\left[ \begin{array}{c} K \\ I \end{array} \right] =
W_0 G
\end{eqnarray}
This implies
\begin{eqnarray} \label{eq:closed_param}
A+BK 
= \left[ \begin{array}{ccc} B & &  A \end{array} \right] W_0 G  
= (X_1-D_0) G
\end{eqnarray}
%\begin{eqnarray} \label{eq:closed_param}
%A+BK &=& \left[ \begin{array}{cc} B & A \end{array} \right]
%\left[ \begin{array}{c} K \\ I \end{array} \right]   \nonumber \\
%&=& \left[ \begin{array}{cc} B & A \end{array} \right] W_0 G  
%= (X_1-D_0) G
%\end{eqnarray}
Thus the formulation \eqref{lqr-form_ideal} coincides with 
the one in \eqref{lqr-form_model}
with $Q=GP$. In particular, $K=U_0QP^{-1}$
and $X_{0} Q = P$ provide an equivalent characterization  
of the two constraints in \eqref{eq:constG}. \\[0.2cm]
Formulation \eqref{lqr-form_ideal} first appeared in \citep{dept2020}
under the assumption that the collected data are noise-free,
that is with $D_0 = 0$, in which case $K_{opt}$ can be directly computed from data.
Here, we revisit this result providing some additional properties 
related to this formulation.

\begin{theorem} \label{thm:main1}
Suppose that condition \eqref{rank.condition} holds. Then 
problem \eqref{lqr-form_ideal} is feasible. Further, 
any optimal solution $(\gamma_o,Q_o,P_o,L_o)$ satisfies 
$K_{opt}=U_0Q_oP_o^{-1}$
%\begin{eqnarray*}
%K_{opt}=U_0Q_oP_o^{-1}
%\end{eqnarray*}
and
\begin{eqnarray*}
\|\mathscr T(K_{opt})\|_2^2 = {\rm trace} ( P_o ) + {\rm trace} ( L_o )
\end{eqnarray*}
\end{theorem} 
The proof of Theorem \ref{thm:main1} relies on two 
auxiliary results. 
\begin{lem} \label{lem:tech1}
Consider any tuple $(\gamma,Q,P,L)$ feasible for \eqref{lqr-form_ideal}.
Then, the controller $K=U_0QP^{-1}$ stabilises \eqref{eq:system} and is such that
\begin{eqnarray*}
\| \mathscr T(K) \|_2^2 \leq {\rm trace}\left( P \right) + {\rm trace} \left( L \right)
\end{eqnarray*}
Proof. \emph{See the appendix}. \qedp
\end{lem}

\begin{lem} \label{lem:tech2}
Suppose that condition \eqref{rank.condition} holds, and 
consider any controller $K$ which stabilises \eqref{eq:system}.
Then, there exists a tuple $(\gamma,Q,P,L)$ feasible for \eqref{lqr-form_ideal}
such that  
$K=U_0QP^{-1}$ and
\begin{eqnarray*}
\| \mathscr T(K) \|_2^2 = {\rm trace}\left( P \right) + {\rm trace} \left( L \right)
\end{eqnarray*}
Proof. \emph{See the appendix}. \qedp
\end{lem}

\emph{Proof of Theorem \ref{thm:main1}}. 
Since $K_{opt}$ is a stabilising controller, Lemma \ref{lem:tech2} ensures the existence
of a tuple $(\gamma,P,Q,L)$ which is
feasible for \eqref{lqr-form_ideal}.
We now consider the second part of the statement.
By Lemma \ref{lem:tech1}, the controller $K = U_0Q_oP_o^{-1}$
satisfies $\|\mathscr T(K)\|^2_2 \leq {\rm trace} (P_o) + {\rm trace} (L_o)$.
Moreover, since $K_{opt}$ is stabilising, by Lemma \ref{lem:tech2}
there  exists a tuple $(\gamma,P,Q,L)$
feasible for \eqref{lqr-form_ideal} such that 
$K_{opt}=U_0Q P^{-1}$ and
$\|\mathscr T(K_{opt})\|_2^2 = {\rm trace} ( P ) + {\rm trace} ( L )$.
As a final step, note that by definition $(\gamma_o,Q_o,P_o,L_o)$ is an optimal
solution to \eqref{lqr-form_ideal}.
Thus ${\rm trace} ( P_o ) + {\rm trace} ( L_o ) \leq {\rm trace} ( P ) + {\rm trace} ( L )$.
In turn, this implies $\|\mathscr T(K)\|_2 \leq \|\mathscr T(K_{opt})\|_2$.
However, since $K_{opt}$ is the controller minimising the $\mathcal H_2$-norm 
of the system we must have
$\|\mathscr T(K)\|_2 = \|\mathscr T(K_{opt})\|_2$
so that $K=K_{opt}$ as the optimal controller is unique. \qedp \\[0.2cm]
As it emerges from the proof of Lemma \ref{lem:tech2},
\eqref{lqr-form_ideal} admits 
infinite optimal solutions in $Q$ of the type $Q=Q_o+Q_\sim$,
where $Q_\sim$ is any matrix in the right kernel of $W_0$.
 
\section{Data-driven solution with noisy data} \label{Sec4}

From previous analysis, when data are noise-free, $K_{opt}$
can be computed directly using \eqref{lqr-form_ideal}. When $D_0 \neq 0$,
\eqref{lqr-form_ideal} cannot be solved unless we know 
$D_0$. In this section, we provide a first solution to 
the case when $D_0$ is nonzero and is not measured. 
This solution offers a quantitative
\emph{relative error} (the gap between the computed solution and $K_{opt}$) 
without making assumptions regarding the noise statistics. \\[0.2cm]
A simple variant of \eqref{lqr-form_ideal} which can be computed 
from data alone and does not involve the knowledge of $D_0$ consists in
disregarding the noise term:
\begin{eqnarray} \label{lqr-form}
\begin{array}{l}
\min_{(\gamma,Q,P,L)} {\gamma}  \\
\textrm{subject to} \smallskip \smallskip \\
\left\{
\def\arraystretch{1.3} 
\begin{array}{l}
X_{1} Q P^{-1} Q^\top X_{1}^\top  - P + I \preceq 0 \\ 
P \succeq I \\
L - U_{0}Q P^{-1} Q^\top U_{0}^\top \succeq 0 \\
X_{0} Q = P \\
 {\rm trace}\left( P \right) + {\rm trace} \left( L \right) \leq \gamma
\end{array} \right.
\end{array}
\end{eqnarray}
If a solution is found then the corresponding controller is computed as
$
K = U_0 Q P^{-1}
$.
Three main questions arise:
\begin{enumerate}[1.]
\item A solution need not exist.
\item Even if a solution is found, the corresponding controller $K$ need to
be stabilising.
\item Even if a solution is found and $K$ is
stabilising, the performance achieved by $K$ might still
substantially differ from the performance achieved by $K_{opt}$.
\end{enumerate}

In the sequel, we will focus on items 2 and 3 above.
Item 1 is implicitly addressed in the analysis. 
Suppose that a solution $(\overline \gamma,\overline Q,\overline P, \overline L)$ 
to \eqref{lqr-form} is found,
and denote by $\overline K = U_0 \overline Q {\overline P}{}^{-1}$ the
corresponding controller. Further, let
$(\gamma_o,Q_o,P_o,L_o)$ be any solution to 
\eqref{lqr-form_ideal} with $K_{opt} = U_0 Q_o P_o^{-1}$.
With this notation in place, we aim at establishing
the following chain of relations:
\begin{eqnarray} \label{eq:chain}
\| \mathscr T(\overline K) \|_2^2
\,\, &\leq& \,\, \eta_1 \left( {\rm trace} ( \overline P ) + {\rm trace} ( \overline L  ) \right) 
\nonumber \\
&\leq& \,\, \eta_1  \eta_2 
\left( {\rm trace} (P_o) + {\rm trace} (L_o) \right) \\
&=& \,\, \eta_1  \eta_2  \| \mathscr T(K_{opt}) \|_2^2 \nonumber
\end{eqnarray}
for some real constants $\eta_1,\eta_2 \geq 1$.
Note in particular that the first inequality ensures that 
$\overline K$ is stabilising. 

\subsection{Stability and performance analysis}

We will focus on the two inequalities in \eqref{eq:chain}
as the equality follows from Theorem \ref{thm:main1}.
Consider the first inequality.
The idea is to find conditions under which  
there exists a constant $\eta_1$ such that 
$\eta_1 (\overline \gamma,\overline Q,\overline P, \overline L)$
is a feasible solution to \eqref{lqr-form_ideal}.
Then the inequality follows from Lemma \ref{lem:tech1}.
For brevity, we introduce some additional
notation. Define
\begin{eqnarray} \label{eq:MTP}
&& M := Q P^{-1} Q^\top \nonumber \\
&& \Theta: = X_{1} M X_{1}^\top  - P \\
&& \Psi: = D_0 M D_0^\top - X_1 M D_0^\top - D_0 M X_1^\top \nonumber
\end{eqnarray}
With this notation the first constraint
in \eqref{lqr-form} reads $\Theta + I \preceq 0$
while the first constraint in \eqref{lqr-form_ideal} reads
$\Theta + \Psi + I \preceq 0$. 
In the sequel, it is understood that all the solutions 
of interest inherit the same notation. In particular,
we will use $\overline M$, $\overline \Theta$ and $\overline \Psi$
to denote the matrices corresponding to 
$(\overline \gamma,\overline Q,\overline P, \overline L)$
and $M_o$, $\Theta_o$ and $\Psi_o$
to denote the matrices corresponding to $(\gamma_o,Q_o, P_o,  L_o)$.

\begin{lem} \label{lem:tech3}
Suppose that \eqref{lqr-form} is feasible. Let 
$(\overline \gamma,\overline Q,\overline P, \overline L)$ be any optimal solution
and let $\overline K = U_0 \overline Q {\overline P}{}^{-1}$.
Let $\eta_1 \geq 1$ be a constant.
If
\begin{eqnarray} \label{eq:constr_1}
\overline \Psi \preceq \left( 1-\frac{1}{\eta_1} \right) I
\end{eqnarray}
Then $\overline K$ 
stabilises system \eqref{eq:system} 
and
\begin{eqnarray}
\| \mathscr T(\overline K) \|_2^2
\leq \eta_1 \left( {\rm trace} ( \overline P ) + {\rm trace} ( \overline L  ) \right)
\end{eqnarray}
\end{lem}

\emph{Proof}. The idea is to show that although
$(\overline \gamma,\overline Q,\overline P, \overline L)$
need not be feasible for \eqref{lqr-form_ideal}, under the condition \eqref{eq:constr_1}
a feasible solution
to \eqref{lqr-form_ideal} is given by $(\hat \gamma, \hat Q,\hat P, \hat L) := 
\eta_1 (\overline \gamma,\overline Q,\overline P, \overline L)$.
Since by hypothesis $(\overline \gamma,\overline Q,\overline P, \overline L)$
is feasible for \eqref{lqr-form}, then 
$(\overline \gamma,\overline Q,\overline P, \overline L)$ satisfies 
$\overline \Theta + I \preceq 0$.
Define $(\hat M,\hat \Theta, \hat \Psi)$ relatice to $(\hat \gamma, \hat Q,\hat P, \hat L)$
as in \eqref{eq:MTP}. 
Then $(\hat M,\hat \Theta, \hat \Psi)  = 
\eta_1 (\overline M,\overline \Theta, \overline \Psi)$
and $\hat \Theta + \eta_1 I= \eta_1 (\overline \Theta + I) \preceq 0$. Thus, 
\begin{eqnarray*}
&& \hat \Theta + \hat \Psi + I  = \\
&& \eta_1 ( \overline \Theta + \overline \Psi ) + \eta_1 I + (1 - \eta_1) I =  \\
&& \eta_1 (\overline \Theta + I) + \eta_1 \overline \Psi + (1 - \eta_1) I \preceq 0
\end{eqnarray*}
where the inequality follows from $\eta_1 (\overline \Theta + I) \preceq 0$ and \eqref{eq:constr_1}.
Hence $(\hat \gamma,\hat Q,\hat P, \hat L)$ satisfies the first constraint
of \eqref{lqr-form_ideal}. Since $(\overline \gamma,\overline Q,\overline P, \overline L)$
is feasible for \eqref{lqr-form} then $(\hat \gamma,\hat Q,\hat P, \hat L)$ 
satisfies by construction 
also all the other constraints of \eqref{lqr-form_ideal}. 
Finally, Lemma \ref{lem:tech1} ensures that
$\overline K = U_0 \overline Q {\overline P}{}^{-1} = 
U_0 \hat Q {\hat P}{}^{-1}$ is stabilising and
\begin{eqnarray*}
\| \mathscr T(\overline K) \|_2^2
\,\, &\leq& \,\,  {\rm trace} ( \hat P ) + {\rm trace} ( \hat L  ) \\
&=& \,\, \eta_1  \left( {\rm trace} ( \overline P ) + {\rm trace} ( \overline L  ) \right) 
\end{eqnarray*}
This gives the claim. \qedp \\[0.2cm]
The second inequality in \eqref{eq:chain} is similar  
to the first one. The point is 
to find conditions under which we can associate to
$K_{opt}$ some tuple $\eta_2 (\gamma_o, Q_o, P_o,  L_o)$ 
feasible for \eqref{lqr-form}.
In this case, the second inequality immediately 
follows from the fact that $(\overline \gamma,\overline Q, \overline P,  \overline L)$ 
is optimal for \eqref{lqr-form}. 

\begin{lem} \label{lem:tech4}
Suppose that condition \eqref{rank.condition} is satisfied, and
let $(\gamma_o,Q_o, P_o,  L_o)$ be any optimal solution 
to \eqref{lqr-form_ideal}.
Let $\eta_2 \geq 1$ be a constant.
If
\begin{eqnarray} \label{eq:constr_2}
- \Psi_o \preceq \left( 1-\frac{1}{\eta_2} \right) I
\end{eqnarray}
then problem \eqref{lqr-form} is feasible. Moreover, any optimal
solution $(\overline \gamma,\overline Q, \overline P, \overline L)$
is such that
\begin{eqnarray} \label{eq:lem:tech4}
{\rm trace} ( \overline P ) + {\rm trace} ( \overline L  ) \leq  
\eta_2 \, \| \mathscr T(K_{opt}) \|_2^2
\end{eqnarray}
\end{lem}

\emph{Proof}. By Theorem \ref{thm:main1} 
condition \eqref{rank.condition} ensures that problem \eqref{lqr-form_ideal}
is feasible and $K_{opt} = U_0 Q_o P_o^{-1}$ where
$(\gamma_o,Q_o, P_o,  L_o)$ is any optimal solution 
to \eqref{lqr-form_ideal}. As before, the idea is to show that although
$(\gamma_o,Q_o, P_o,  L_o)$
need not be feasible for \eqref{lqr-form}, under condition \eqref{eq:constr_2} a feasible solution
to \eqref{lqr-form} is given by $(\tilde \gamma, \tilde Q,\tilde P, \tilde L)  
:= \eta_2 (\gamma_o,Q_o, P_o,  L_o)$.
Since $(\gamma_o,Q_o, P_o,  L_o)$
is feasible for \eqref{lqr-form_ideal} then 
$(\gamma_o,Q_o, P_o,  L_o)$ satisfies 
$\Theta_o + \Psi_o + I \preceq 0$.
Define $(\tilde M, \tilde \Theta,\tilde \Psi)$ relative to $(\tilde \gamma, \tilde Q,\tilde P, \tilde L)$
as in \eqref{eq:MTP}. Then
$(\tilde M, \tilde \Theta,\tilde \Psi) = \eta_2 (M_o, \Theta_o,  \Psi_o)$
and 
$\tilde \Theta + \tilde \Psi + \eta_2 I= \eta_2 (\Theta_o + \Psi_o + I) \preceq 0$.
%\begin{eqnarray*}
%\tilde \Theta + \tilde \Psi + \eta_2 I= \eta_2 (\Theta_o + \Psi_o + I) \preceq 0
%\end{eqnarray*}
Hence, 
\begin{eqnarray*}
&& \tilde \Theta + I  = \\
&& \eta_2 (  \Theta_o + \Psi_o ) - \eta_2 \Psi_o  + \eta_2 I + (1 - \eta_2) I =  \\
&& \eta_2 (  \Theta_o + \Psi_o + I) - \eta_2 \Psi_o + (1 - \eta_2) I \preceq 0
\end{eqnarray*}
where the inequality follows from $\eta_2 (  \Theta_o + \Psi_o + I) \preceq 0$ 
and \eqref{eq:constr_2}.
Hence $(\tilde \gamma,\tilde Q,\tilde P, \tilde L)$ satisfies the first constraint
of \eqref{lqr-form}. Since $(\gamma_o,Q_o, P_o,  L_o)$
is feasible for \eqref{lqr-form_ideal} then $(\tilde \gamma,\tilde Q,\tilde P, \tilde L)$ 
satisfies by construction 
also all the other constraints of \eqref{lqr-form}. Hence the claim
follows since $(\overline \gamma,\overline Q, \overline P, \overline L)$ is optimal
and since the cost associated with $(\tilde \gamma,\tilde Q,\tilde P, \tilde L)$ 
satisfies
\begin{eqnarray*}
{\rm trace} ( \tilde P ) + {\rm trace} ( \tilde L  ) &=&  
\eta_2 ( {\rm trace} ( P_o ) + {\rm trace} ( L_o ) ) \\
&=& \eta_2 \, \| \mathscr T(K_{opt}) \|_2^2
\end{eqnarray*}
which establishes \eqref{eq:lem:tech4}.
\qedp

%By combining these two lemmas the following result can
%be stated. 

%\begin{theorem}  \label{thm:main2}
%Suppose that \eqref{rank.condition} holds.
%Under \eqref{eq:constr_1} and \eqref{eq:constr_2},
%a solution to \eqref{lqr-form} exists. Moreover, the resulting controller
%$\overline K$ is stabilising and satisfies 
%$\| \mathscr T(\overline K) \|_2^2 \leq \eta_1 \eta_2 \| \mathscr T(K_{opt}) \|_2^2$.
%\qedp
%\end{theorem}

We summarize our findings in the following result.

\begin{theorem}  \label{thm:main2}
Let $U_0$, $X_0$ and $X_1$ be data generated from
an experiment on system \eqref{eq:system} possibly with 
nonzero disturbance vector $D_0$.
Consider problem \eqref{lqr-form}. 
\begin{enumerate} 
\item[(i)] Under condition \eqref{eq:constr_1}, 
any optimal solution $(\overline \gamma,\overline Q,\overline P, \overline L)$ 
to \eqref{lqr-form} is such that  
the controller $\overline K = U_0 \overline Q {\overline P}{}^{-1}$ 
stabilizes \eqref{eq:system} with guaranteed performance 
\begin{eqnarray*} 
\| \mathscr T(\overline K) \|_2^2
\leq \eta_1 \left( {\rm trace} ( \overline P ) + {\rm trace} ( \overline L  ) \right)
\end{eqnarray*} 
%$\| \mathscr T(\overline K) \|_2^2
%\leq \eta_1 \left( {\rm trace} ( \overline P ) + {\rm trace} ( \overline L  ) \right)$,
with $\eta_1$ as in \eqref{eq:constr_1}.
\item[(ii)] If, in addition to \eqref{eq:constr_1}, 
also the conditions \eqref{rank.condition} and \eqref{eq:constr_2} hold then
\eqref{lqr-form} is feasible and 
the controller $\overline K$ satisfies
$\| \mathscr T(\overline K) \|_2^2 \leq \eta_1 \eta_2 \| \mathscr T(K_{opt}) \|_2^2$,
or equivalently, 
\begin{eqnarray*} 
\frac{\| \mathscr T(\overline K) \|_2^2 - \| \mathscr T(K_{opt}) \|_2^2}
{\| \mathscr T(K_{opt}) \|_2^2}
\leq ( \eta_1 \eta_2 -1 ) 
\end{eqnarray*}
with $\eta_2$ as in \eqref{eq:constr_2}. \qedp
\end{enumerate}
\end{theorem}

\subsection{Preliminary discussion}
 
The bound in (ii) of the above theorem defines the relative error with respect 
to the optimal solution. In our setting, this bound 
holds with no prior assumptions on the noise statistics, for instance 
the noise being a martingale or white. We note in particular that 
this error nicely scales with $\eta_1$ and $\eta_2$, and converges
to zero as $D_0$ goes to zero since, in this case, both $\eta_1$ and $\eta_2$ 
converge to one. \\[0.2cm]
Conditions \eqref{eq:constr_1} and \eqref{eq:constr_2} play 
a different role. The first one
ensures that \emph{any} solution to problem \eqref{lqr-form} returns a 
stabilizing controller. This condition 
can be checked from data alone whenever some prior 
information on $d$ is available. Instead, 
condition \eqref{eq:constr_2}
makes it possible to explicitly 
quantify the performance gap between the solution and $K_{opt}$.
Different from \eqref{eq:constr_1}, condition 
\eqref{eq:constr_2} cannot be checked from data as it depends on the (unknown)
optimal controller $K_{opt}$. In the next section, we will 
nonetheless discuss an interesting fact related to \eqref{eq:constr_2},
namely that this condition is actually easier to 
satisfy in practice than \eqref{eq:constr_1}. 
We postpone a discussion on this point to Section \ref{Sec5} and first
consider a variant of \eqref{lqr-form} with the goal 
of rendering \eqref{eq:constr_1} easier to fulfil.

\section{Noise robustness through soft constraints} \label{Sec5}

Condition \eqref{eq:constr_1} may be difficult to satisfy unless $d$ has very small 
magnitude. In fact, in order to satisfy \eqref{eq:constr_1} one needs
$\overline \Psi \prec I$, where
\begin{eqnarray} \label{}
\overline \Psi: = D_0 \overline M D_0^\top - 
X_1 \overline M D_0^\top - D_0 \overline M X_1^\top
\end{eqnarray}
with $\overline M = \overline Q {\overline P}{}^{-1} \overline Q^\top$.
However, there is no constraint on the magnitude of $\overline M$ in \eqref{lqr-form}
with the consequence that a small level of noise 
may generate non-stabilizing controllers.
This observation hints at modifying \eqref{lqr-form} by 
adding a constraint on
the magnitude of $M= Q P^{-1} Q^\top$. 
Consider the SDP:
\begin{eqnarray} \label{lqr-forma}
\begin{array}{l}
\min_{(\gamma,Q,P,L,V)} {\gamma}  \\
\textrm{subject to} \smallskip \smallskip \\
\left\{
\def\arraystretch{1.3} 
\begin{array}{l}
X_{1} Q P^{-1} Q^\top X_{1}^\top  - P + I \preceq 0 \\ 
P \succeq I \\
L - U_{0}Q P^{-1} Q^\top U_{0}^\top \succeq 0 \\
V - Q P^{-1} Q^\top \succeq 0 \\
X_{0} Q = P \\
 {\rm trace}\left( P \right) + {\rm trace} \left( L \right) + {\rm trace} \left( V \right) \leq \gamma
\end{array} \right.
\end{array}
\end{eqnarray}
Compared with \eqref{lqr-form}, we now search for solutions 
that lead to matrices $M=Q P^{-1} Q^\top$ having small trace,
equivalently such that $Q P^{-\frac{1}{2}}$ has
small {singular values}. A \emph{soft} constraint 
favours small values of $M$ while preserving all the logical
steps of the baseline solution.
\\[0.2cm]
We proceed as before.
Assume that a solution $(\overline \gamma,\overline Q,\overline P, \overline L, \overline V)$ 
to \eqref{lqr-forma} is found
and let $\overline K = U_0 \overline Q {\overline P}{}^{-1}$ be the
corresponding controller. Let
$(\gamma_o,Q_o,P_o,L_o)$ be any optimal solution to 
\eqref{lqr-form_ideal} with $K_{opt} = U_0 Q_o P_o^{-1}$.
We aim at establishing
the following chain of relations:
\begin{eqnarray} \label{eq:chain2}
\| \mathscr T(\overline K) \|_2^2 
\,\, &\leq& \,\, \eta_1 \left( {\rm trace} ( \overline P ) + {\rm trace} ( \overline L  )  \right) \nonumber \\
&\leq& \,\, \eta_1  \eta_2 
\left( {\rm trace} ( P_o ) + {\rm trace} ( L_o  ) + {\rm trace} ( V_o  ) \right) \nonumber \\
&=& \,\, \eta_1  \eta_2  \| \mathscr T(K_{opt}) \|_2^2
+  \eta_1 \eta_2 {\rm trace} ( V_o  ) 
\end{eqnarray}
for some real constants $\eta_1,\eta_2 \geq 1$
with 
\begin{eqnarray} \label{eq:Vo}
V_o := Q_o P_o^{-1} Q_o^\top = M_o
\end{eqnarray}

\subsection{Stability and performance analysis}

The first inequality follows as in
Lemma \ref{lem:tech3}.
\begin{lem} \label{lem:tech5}
Suppose that \eqref{lqr-forma} is feasible. Let
$(\overline \gamma, \overline Q,\overline P, \overline L, \overline V)$ be
any optimal solution
and let $\overline K = U_0 \overline Q {\overline P}{}^{-1}$.
Let $\eta_1 \geq 1$ be a constant.
If condition \eqref{eq:constr_1} is satisfied 
then $\overline K$ stabilises system \eqref{eq:system} 
and
\begin{eqnarray}
\| \mathscr T(\overline K) \|_2^2 
\leq \eta_1 \left( {\rm trace} ( \overline P ) + {\rm trace} ( \overline L  ) \right)
\end{eqnarray}
\end{lem}

\emph{Proof}. The proof is analogous to the one of 
Lemma \ref{lem:tech3} and therefore omitted. (Note in particular
that the constraint in \eqref{lqr-forma} that involves $V$
does not appear in \eqref{lqr-form_ideal}.)
\qedp

We also have a natural counterpart of Lemma \ref{lem:tech4},
which establishes the second inequality in \eqref{eq:chain2}.

\begin{lem} \label{lem:tech6}
Suppose that condition \eqref{rank.condition} is satisfied, and
let $(\gamma_o,Q_o, P_o,  L_o)$ be any optimal solution 
to \eqref{lqr-form_ideal}.
Let $\eta_2 \geq 1$ be a constant.
If condition \eqref{eq:constr_2} is satisfied 
then problem \eqref{lqr-forma} is feasible. Moreover, any optimal
solution $(\overline \gamma,\overline Q, \overline P, \overline L, \overline V)$
is such that
\begin{eqnarray} \label{eq:lem:tech6}
&& {\rm trace} ( \overline P ) + {\rm trace} ( \overline L  ) \leq  \nonumber \\
&& \qquad \qquad \eta_2 \left( \| \mathscr T(K_{opt}) \|_2^2 + {\rm trace} ( V_o ) \right)
\end{eqnarray}
\end{lem}

\emph{Proof}. The proof is essentially analogous to the proof of 
Lemma \ref{lem:tech4}. By
proceeding as in Lemma \ref{lem:tech4} it is immediate to 
verify that $\eta_2 (\gamma_o,Q_o, P_o,  L_o)$ is a feasible 
solution to \eqref{lqr-forma} where 
the constraint in \eqref{lqr-forma} involving $V$
is satisfied by the choice
$V_o = Q_o P_o^{-1} Q_o^\top$. This implies
\begin{eqnarray*}
&& {\rm trace} ( \overline P ) + {\rm trace} ( \overline L  ) + {\rm trace} ( \overline V  )  \leq  \\
&& \qquad \qquad \eta_2 \left( {\rm trace} ( P_o ) + {\rm trace} ( L_o  )  
+ {\rm trace} ( V_o  )  \right) 
\end{eqnarray*}
which establishes \eqref{eq:lem:tech6}. 
\qedp \\[0.2cm]
By combining these two lemmas the following result can
be stated. 

\begin{theorem}  \label{thm:main3}
Let $U_0$, $X_0$ and $X_1$ be data generated from
an experiment on system \eqref{eq:system} possibly with 
nonzero disturbance vector $D_0$.
Consider problem \eqref{lqr-forma}. Then:
\begin{enumerate} 
\item[(i)] Under condition \eqref{eq:constr_1}, 
any optimal solution $(\overline \gamma,\overline Q,\overline P, \overline L)$ 
to \eqref{lqr-forma} is such that
the controller $\overline K = U_0 \overline Q {\overline P}{}^{-1}$  stabilizes \eqref{eq:system} 
with guaranteed performance 
\begin{eqnarray*} 
\| \mathscr T(\overline K) \|_2^2
\leq \eta_1 \left( {\rm trace} ( \overline P ) + {\rm trace} ( \overline L  ) \right)
\end{eqnarray*} 
with $\eta_1$ as in \eqref{eq:constr_1}.
\item[(ii)] If, in addition to \eqref{eq:constr_1}, 
also the conditions \eqref{rank.condition} and \eqref{eq:constr_2} hold then
\eqref{lqr-forma} is feasible and 
the controller $\overline K$ satisfies
$\| \mathscr T(\overline K) \|_2^2 \leq \eta_1 \eta_2 \| \mathscr T(K_{opt}) \|_2^2
+ \eta_1 \eta_2 {\rm trace} ( V_o  )$,
or equivalently, 
\begin{eqnarray*} 
\frac{\| \mathscr T(\overline K) \|_2^2 - \| \mathscr T(K_{opt}) \|_2^2}
{\| \mathscr T(K_{opt}) \|_2^2} 
\leq ( \eta_1 \eta_2 -1 ) + \eta_3
\end{eqnarray*}
with $\eta_2$ as in \eqref{eq:constr_2}, and where
\begin{eqnarray*} 
\eta_3 := \eta_1 \eta_2 \frac{{\rm trace} ( V_o  )}{\| \mathscr T(K_{opt}) \|_2^2}
\end{eqnarray*}
with $V_o$ as in \eqref{eq:Vo}. 
\qedp
\end{enumerate}
\end{theorem}

Compared to the baseline solution, the error bound 
worsen due to the extra term $\eta_3$.
This matches what is observed in indirect methods  \citep{Mania2019}
for noise-robust solutions.
As we discuss in Section \ref{Sec6}, the conservatism can be nonetheless 
kept to moderate values. 

\begin{rem} {\rm (Implementation of \eqref{lqr-forma})}
Problem \eqref{lqr-forma} can be given the equivalent form of an SDP:
\begin{eqnarray} \label{lqr-form-explicit-sdp}
\begin{array}{l}
\min_{(\gamma,Q,P,L)} {\gamma}  \\
\textrm{subject to} \smallskip \smallskip \\
\left\{
\def\arraystretch{1.3} 
\begin{array}{l}
\begin{bmatrix}
I - P & X_{1} Q \\
Q^\top X_{1}^\top & -P
\end{bmatrix} \preceq 0 \\[0.4cm]
\begin{bmatrix}
L & U_0 Q\\
Q^\top U_0^\top  & P
\end{bmatrix}
\succeq 0 \\[0.4cm]
\begin{bmatrix}
V & Q\\
Q^\top & P
\end{bmatrix} \succeq 0 \\[0.4cm]
X_{0} Q = P \\
 {\rm trace}\left( P \right) + {\rm trace} \left( L \right) + {\rm trace} \left( V \right) \leq \gamma
\end{array} \right.
\end{array}
\end{eqnarray}
Similar considerations apply to \eqref{lqr-form}. \qedp
\end{rem}

\subsection{Alternative based on the $S$-procedure} \label{subsec:robustified}

The key to overcome the lack of knowledge about $D_0$ 
in the problem \eqref{lqr-form_ideal}  is to completely disregard 
such a term to obtain the implementable form \eqref{lqr-form}, 
from which a robust version with soft constraints on $M$ \eqref{lqr-forma} 
is eventually derived.  An alternative to this approach is to explicitly impose 
conditions on $D_0$ with the perspective of obtaining a 
potentially more robust version  of \eqref{lqr-form_ideal} 
and \eqref{lqr-forma}. We discuss advantages and drawbacks of the approach. \\[0.2cm]
As in \citep{dept2020}, we initially consider a condition on $D_0$ in the form 
\begin{eqnarray}\label{snr}
D_0 D_0^\top \preceq \mu^2 RR^\top
\end{eqnarray}   
where $R$ is a full-row rank matrix and 
$\mu$ is a positive real. 
In the remainder of this subsection, 
we revisit this condition with the purpose of  providing a variant of  \eqref{lqr-forma}. 
We let the constraint \eqref{snr} be \emph{regular}, 
assuming that there exists a vector $\overline x$ such that 
$\overline x^\top D_0 D_0^\top \overline x < \mu^2 \overline x^\top  
RR^\top \overline x$. This condition is only introduced to motivate the formulation of the more robust variant of  \eqref{lqr-forma} and will be not be required in the main result of the section.
\\[0.2cm]
Under \eqref{snr}, by the $S$-procedure \citep[Section 2.1.2]{kh2004stability},  
a necessary and sufficient condition to ensure the robust 
stabilizability condition in \eqref{lqr-form_ideal}, namely to ensure that  
\begin{eqnarray*}
(X_{1} -D_0) Q P^{-1} Q^\top (X_{1}-D_0)^\top  - P +I \preceq 0 
\end{eqnarray*}
is the existence of a parameter $\tau\ge 0$ such that  
\begin{eqnarray*}
\begin{array}{l}
(X_{1} -D_0) Q P^{-1} Q^\top (X_{1}-D_0)^\top  - P +I \\
\hspace{3.5cm}-\tau D_0 D_0^\top + \mu^2 \tau RR^\top \preceq 0
\end{array}
\end{eqnarray*}
To get rid of the dependence on the unknown matrix $D_0$, 
instead of the equivalent condition above, one can consider the sufficient condition 
\begin{eqnarray}\label{robust.lmi}
\begin{bmatrix}
- P + X_{1} M X_{1}^\top +\mu^2 \tau RR^\top + I & - X_1 M \\
- M X_{1}^\top  & M- \tau I \\
\end{bmatrix}\!
\preceq 0
\end{eqnarray}
for some $\tau\ge 0$, with $M$ as in \eqref{eq:MTP}, 
and include this stronger variation in the following alternative 
to problem \eqref{lqr-form}:
\begin{eqnarray} \label{lqr-form-alternative}
\begin{array}{l}
\min_{(\gamma,Q,P,L, \tau)} {\gamma}  \\
\textrm{subject to} \smallskip \smallskip \\
\left\{
\def\arraystretch{1.3} 
\begin{array}{l}
\begin{bmatrix}
- P + X_{1} M X_{1}^\top +\mu^2 \tau RR^\top +\frac{1}{\eta_1} I  & - X_1 M \\
- M X_{1}^\top  & M- \tau I\\
\end{bmatrix}\!\!\!
\preceq 0
\\ 
P \succeq I \\
L - U_{0}Q P^{-1} Q^\top U_{0}^\top \succeq 0 \\
X_{0} Q = P \\
 {\rm trace}\left( P \right) + {\rm trace} \left( L \right) \leq \gamma\\
\tau\ge 0
\end{array} \right.
\end{array}
\end{eqnarray}
where the role of the factor $\eta_1\geq1$ 
will be explained later. 
We have previously discussed on the 
benefits of including a soft constraint on $M=QP^{-1} Q^\top$. 
We would like to have these benefits also in the robust variant 
of the SDP problem that we are studying here. We recall that the 
idea consists of bounding $M$ via a matrix decision 
variable $V$, whose trace is made as small as possible in the 
optimization process. This is achieved by introducing the new 
constraint $V-QP^{-1} Q^\top\succeq 0$ and by modifying the 
existing constraint ${\rm trace}\left( P \right) + {\rm trace} \left( L \right) \leq \gamma$ 
into ${\rm trace}\left( P \right) + {\rm trace} \left( L \right) + {\rm trace} \left( V \right) \leq \gamma$. \\[0.2cm]
Let us observe now that a constraint on $M$ is already 
present in the first constraint of \eqref{lqr-form-alternative} in the form 
$M- \tau I \preceq 0$. Thus, it is natural to modify block $(2,2)$ 
in that constraint as $M- V \preceq 0$. However, the matrix 
$-\tau I$ in $M- \tau I \preceq 0$ derived from imposing the 
condition \eqref{snr} and the latter also motivated the introduction 
of the term $\mu^2 \tau RR^\top$ in block $(1,1)$ in the first 
constraint of \eqref{lqr-form-alternative}, which must be changed accordingly. 
These arguments lead to the following modification of 
\eqref{lqr-form-alternative} : 
\begin{eqnarray} \label{lqr-formb}
\begin{array}{l}
\min_{(\gamma,Q,P,L,V)} {\gamma}  \\
\textrm{subject to} \smallskip \smallskip \\
\left\{
\def\arraystretch{1.3} 
\begin{array}{l}
\begin{bmatrix}
- P + X_{1} M X_{1}^\top +\mu^2 R V R^\top +\frac{1}{\eta_1} I& - X_1 M \\
- M X_{1}^\top  & M-V\
\end{bmatrix}\preceq 0
\\ 
P \succeq I \\
L - U_{0}Q P^{-1} Q^\top U_{0}^\top \succeq 0 \\
X_{0} Q = P \\
 {\rm trace}\left( P \right) + {\rm trace} \left( L \right) + {\rm trace} \left( V \right)\leq \gamma
\end{array} \right.
\end{array}
\end{eqnarray}
We now establish the main properties of \eqref{lqr-formb}.

\begin{theorem}  \label{thm:main4}
Let $U_0$, $X_0$ and $X_1$ be data generated from
an experiment on system \eqref{eq:system} possibly with 
nonzero disturbance vector $D_0$.
Assume that problem \eqref{lqr-formb} is feasible with $\eta_1 \geq 1$. 
Let $(\overline \gamma, \overline Q, \overline P, \overline L, \overline V)$ be any optimal solution 
to \eqref{lqr-formb} and let $\overline K=U_0 \overline Q\, \overline P^{-1}$.
If
\begin{eqnarray} \label{snr.V}
D_0 \overline V D_0^\top \preceq \mu^2 R \overline V R^\top
\end{eqnarray}
then $\overline K$ stabilises system 
\eqref{eq:system} 
and 
\begin{eqnarray*}
\|\mathscr{T}(\overline K)\|_2^2 \le \eta_1 ({\rm trace}(P)+{\rm trace}(L))
\end{eqnarray*}
\end{theorem}
\emph{Proof}.  Similarly to the proof of Lemma \ref{lem:tech3}, the 
idea is to let 
$(\hat \gamma, \hat Q, \hat P, \hat L):=
\eta_1(\overline \gamma, \overline Q, \overline P, \overline L)$
and show that it satisfies the first constraint in \eqref{lqr-form_ideal}. 
In fact, by multiplying the first constraint in \eqref{lqr-formb}
by the vector $x^\top \left[I \,\,\,D_0\right]$ on the left and its transpose 
on the right, it is straightforward to see that \eqref{lqr-formb}  implies 
\begin{eqnarray*}
\begin{array}{r}
-\overline P + X_1 \overline M  X_{1}^\top+\mu^2 R \overline V R^\top +\frac{1}{\eta_1} I- X_1  \overline M D_0^\top\\
- D_0  \overline M  X_1^\top+D_0  \overline M D_0^\top- 
D_0 \overline V D_0^\top\preceq 0
\end{array}
\end{eqnarray*}
Hence, in view of \eqref{snr.V},  
\begin{eqnarray}\label{namely.we.show}
(X_{1} -D_0) \hat Q \hat P^{-1} \hat Q^\top (X_{1}-D_0)^\top  -\hat  P + I \preceq 0
\end{eqnarray}
\emph{i.e.}, it satisfies the first constraint in \eqref{lqr-form_ideal}.
Similar to what has been noticed in the proof of Lemma \ref{lem:tech5}, 
by construction and thanks to the condition $\eta_1\ge 1$,  the solution 
$(\hat \gamma, \hat Q, \hat P, \hat L)$ satisfies all the other constraints 
in \eqref{lqr-form_ideal} and the  thesis follows from Lemma \ref{lem:tech1}. 
\qedp

This alternative formulation relies on arguments typical of the
$S$-procedure for robust control, where here the noise is regarded
as the source of uncertainty. 
Compared with \eqref{lqr-forma}, the formulation \eqref{lqr-formb} has the 
advantage to \emph{explicitly} address robust stabilization 
by incorporating $\eta_1$ directly into design. 
We see that this parameter embodies a tradeoff: 
the smaller its value is the higher is the chance to satisfy the first constraint 
in \eqref{lqr-formb}, hence to have a stabilizer robust to noise, 
but the coarser is the upper bound on the $\mathcal H_2$-norm 
of the closed-loop system. 
On the other hand, the drawback of this formulation 
is that it is not clear at the moment how to establish the analogous 
of Lemma \ref{lem:tech6} and therefore of Theorem \ref{thm:main3}, 
which give us a quantitative bound on the error function. 
Overall, the expectation
is that \eqref{lqr-formb} provides
an increased robustness to noise than \eqref{lqr-forma} at the 
expense of decreased performance 
in terms of optimality, similarly to what we have when comparing \eqref{lqr-forma}
to \eqref{lqr-form}. This
expectation is confirmed by the Monte Carlo 
simulations that we perform in Section \ref{Sec7}. 

\begin{rem}
{\rm (Implementation of \eqref{lqr-formb})}
Since $X_0 Q=P$ and $M=QP^{-1} Q^\top$, by Schur complement 
the first inequality in  \eqref{lqr-formb} becomes
\begin{eqnarray*}
\begin{bmatrix}
- X_0 Q +\mu^2  RV R^\top +\frac{1}{\eta_1} I& 0 & X_1 Q \\
0 & - V & - Q\\
(X_1Q)^\top & -Q^\top & - X_0 Q
\end{bmatrix}
\preceq 0 
\end{eqnarray*}
which gives rise to an SDP analogous to \eqref{lqr-form-explicit-sdp}. 
Problem \eqref{lqr-formb} can be then implemented through a line 
search on $\eta_1$ as we will illustrate in the numerical simulations.
\qedp
\end{rem}

\section{Stability and performance verification, 
nonlinear systems and de-noising} \label{Sec6}

We devote this section to discuss some practical aspects of the
proposed method as well as possible extensions of the previous 
analysis.

\subsection{Stability and performance verification}

Sections \ref{Sec5} and  \ref{Sec6} give stability 
and performance properties of our data-driven approach to the 
LQR problem. We now discuss how to
infer these properties by only looking at the data.

\subsubsection{Stability and $\mathcal H_2$-norm bounds}

Inferring stability and performance of $\overline K$
inevitably requires some prior assumptions
on the quality of data, hence on the noise. Our method makes a
parsimonious use of such priors (\emph{i.e.} no noise statistics are needed),
much in the spirit of robust control design \citep{scherer-LMI-book}. \\[0.2cm]
To ensure stability with an $\mathcal H_2$-norm bound,
Theorem \ref{thm:main2} and \ref{thm:main3} require the fulfilment 
of the condition \eqref{eq:constr_1} which involves the
noise-dependent matrix 
\begin{eqnarray*}
\overline \Psi = D_0 \overline M D_0^\top - X_1 \overline M D_0^\top - D_0 \overline 
M X_1^\top
\end{eqnarray*}
If one knows that $\|D_0\| \leq \delta$ for some $\delta >0$
then condition \eqref{eq:constr_1} is satisfied if
\begin{eqnarray} \label{eq:stab_practical}
\delta^2 \|\overline M\| + 2 \delta \|X_1 \overline M\| \leq 
 1 - \displaystyle\frac{1}{\eta_1} 
\end{eqnarray}
for some $\eta_1 \geq 1$,
which can be checked from data alone. 
A bound on $\|D_0\|$ can be obtained from prior 
information on the noise magnitude, and is representative 
of the noise energy content. 
Condition \eqref{eq:stab_practical}
has the same features of the original condition \eqref{eq:constr_1}
in the sense that is becomes easier to satisfy
for small values of $\overline M$. \\[0.2cm]
%We observe in particular that  
%\eqref{eq:closed_param}
%implies
%\begin{eqnarray*} \label{}
%X_1 M = (A+B K) Q^\top + D_0 M
%\end{eqnarray*}
%This shows that condition \eqref{eq:constr_1} (hence \eqref{eq:stab_practical}) can be
%satisfied even when $X_1$ takes on large values
%provided that $Q$ is small. 
%Numerical simulations indicate that this is
%what happens in general 
%because $M=QP^{-1}Q^\top$ 
%and because large values of $P$ imply 
%large values of the cost, so that $M$
%is made small through $Q$. \\[0.2cm]
%
Similar considerations apply to Theorem \ref{thm:main4} where,
under the bound $\|D_0\| \leq \delta$, condition \eqref{snr.V}
is satisfied if
\begin{eqnarray} \label{eq:stab_practicala}
\delta^2 \| \overline V \| I \preceq \mu^2 R \overline V R^\top
\end{eqnarray}
which can be checked from data alone. We note that 
for both \eqref{eq:stab_practical} and \eqref{eq:stab_practicala}
there is an evident tradeoff between priors and conservativeness:
the larger the value of $\delta$ is the higher is the chance that the assumption
on the noise is satisfied, but the lower is the chance that 
\eqref{eq:stab_practical} and \eqref{eq:stab_practicala} hold.
We also note that approaches other than the one discussed 
can be used. In particular, if $D_0$ is known to belong to some {compact} set 
$\mathscr D$ then one can check \eqref{eq:constr_1} through a finite set of linear matrix 
inequalities computed at the vertices of a {convex} embedding of $\mathscr D$
as done in \citep{Bisoffi2019}, albeit this approach is computationally demanding. 

\subsubsection{Bounds on the relative error} \label{Subsec_error}

To achieve bounds on the error also conditions \eqref{rank.condition} and 
\eqref{eq:constr_2} are used.
Condition \eqref{rank.condition} 
states that the data are sufficiently rich in content.
As noted in \citep{Waarde2020} 
this condition is not restrictive for the LQR problem.
In fact, in the noiseless case, this condition is necessary for reconstructing 
$A$ and $B$ from data (hence for any {model-based} solution
and indirect data-driven method)
and is also generically necessary for any direct {data-driven} method.\footnote{Condition
\eqref{rank.condition} is instead not needed in general to find a stabilizing controller
(\emph{cf.} Theorems \ref{thm:main2}, \ref{thm:main3} and \ref{thm:main4}). This matches 
the observations made in \citep{Waarde2020}.}\\[0.2cm]
Condition \eqref{rank.condition} can be checked 
from data. In the noise-free case it can be enforced 
at the experiment stage (Lemma \ref{lem:willems}). In the noisy case
\eqref{rank.condition} may or may not hold depending 
on the noise level (although it always remains checkable from data).
Nonetheless, it is simple to see that \eqref{rank.condition} continues
to hold in the presence of noise when $d$ is sufficiently small. 
In fact, by linearity, $W_0$ can be decomposed as 
\begin{eqnarray*} 
W_0 = \left[ 
\begin{array}{c} U_0 \\ X_0^u \end{array}
\right] + \left[ 
\begin{array}{c} 0 \\ X^d_0 \end{array}
\right]
\end{eqnarray*}
where $X^u_0$ and $X^d_0$ represent the state data generated by $u$
and $d$, respectively. Since by Lemma \ref{lem:willems} the matrix involving 
$U_0$ and $X_0^u$ has full row rank then also $W_0$ has full row rank
whenever $d$ has sufficiently low magnitude. \\[0.2cm]
We next focus on the condition \eqref{eq:constr_2}.
The structure of \eqref{eq:constr_2} is analogous to \eqref{eq:constr_1},
with the difference that it involves the matrix
$\Psi_o = D_0 M_o D_0^\top - X_1 M_o D_0^\top - D_0 M_o X_1^\top$
%\begin{eqnarray*}
%\Psi_o = D_0 M_o D_0^\top - X_1 M_o D_0^\top - D_0 M_o X_1^\top
%\end{eqnarray*}
instead of $\overline \Psi$. Like condition \eqref{eq:constr_1},
also \eqref{eq:constr_2} is
automatically satisfied for $D_0=0$, in which case $\eta_2 =1$. 
Different from \eqref{eq:constr_1}, \eqref{eq:constr_2}
cannot be checked from data as it depends on the (unknown)
optimal controller $K_{opt}$ via $M_o$. 
Nonetheless, an interesting fact related to \eqref{eq:constr_2}
is that this condition is actually easier to satisfy than  \eqref{eq:constr_1}. 
This indicates in particular that the robust solution in Theorem \ref{thm:main3}
does not introduce much conservatism with respect to the baseline solution
in Theorem \ref{thm:main2}.
We now elaborate on this point. \\[0.2cm]
In both Theorems \ref{thm:main2} and \ref{thm:main3}, the 
performance gap between $\overline K$ and $K_{opt}$ holds 
for \emph{any} optimal solution $(\gamma_o,Q_o,P_o,L_o)$ 
to problem \eqref{lqr-form_ideal}, and this is possible since by Theorem \ref{thm:main1} 
all the solutions are such that $K_{opt}=U_0Q_oP_o^{-1}$
with $\mathcal H_2$-norm
$\|\mathscr T(K_{opt})\|_2^2 = {\rm trace} ( P_o ) + {\rm trace} ( L_o )$.
We now derive a particular (optimal) solution,
the derivation being analogous to the one in Lemma \ref{lem:tech2}.
Let $P_o \succ 0$ be the unique solution to
$(A+B K_{opt}) P_o (A+B K_{opt})^\top  - P_o + I = 0$.
In particular, $P_o$ is the \emph{controllability Gramian}
of the closed-loop system. Let now
\begin{eqnarray} \label{eq:Go_special}
G_o := 
W_0^\dag 
\left[ \begin{array}{c}
 K_{opt} \\ I 
\end{array} \right]
\end{eqnarray}
Finally, define $Q_o := G_o P_o$,
$L_o := U_0 Q_o P_o^{-1} Q_o^\top U_0^\top$ and
$\gamma_o:={\rm trace} \left( P_o \right) + {\rm trace} \left( L_o \right)$.
By definition of $\mathcal H_2$-norm,
\begin{eqnarray*}
\|\mathscr T( K_{opt})\|_2^2 \,\,
&=&\,\, {\rm trace} \left( P_o \right) + {\rm trace} \left( K_{opt}^\top P_o K_{opt} \right) \\
&=&\,\, {\rm trace} \left( P_o \right) + {\rm trace} 
\left( U_0 Q_o P_o^{-1} Q_o^\top U_0^\top \right) \\
&=&\,\, {\rm trace} \left( P_o \right) + {\rm trace} \left( L_o \right) 
\end{eqnarray*}
This particular solution is optimal as it achieves the same
cost of any other optimal solution (Theorem \ref{thm:main1}). 
The special feature of this
solution is that $G_o$ 
is the \emph{minimum norm} least-squares solution to 
\eqref{eq:constG} with $K=K_{opt}$,
and so is $Q_o = G_o P_o$. 
Since $M_o=Q_o P_o^{-1} Q_o^\top$, condition
\eqref{eq:constr_2} turns out to be satisfied more easily
than \eqref{eq:constr_1} since the matrix $\overline M$
appearing in \eqref{eq:constr_1} is instead not necessarily associated 
to any minimum norm solution. We note that
$Q_o$ (thus $M_o$) decreases as the norm of $W_0$ increases, 
which happens for instance when the number $T$ of collected data
increases. This implies in particular that $V_o$ approaches $0$ as $W_0$ increases.
In turn, this means that the formulation \eqref{lqr-forma} 
does not introduce much conservatism with respect to 
the formulation \eqref{lqr-form} 
since the performance bound
\begin{eqnarray*} 
\| \mathscr T(\overline K) \|_2^2 \leq \eta_1 \eta_2 \| \mathscr T(K_{opt}) \|_2^2 
+ \eta_1 \eta_2 {\rm trace} ( V_o ) 
\end{eqnarray*} 
approaches the bound  
$\| \mathscr T(\overline K) \|_2^2 \leq \eta_1 \eta_2 \| \mathscr T(K_{opt}) \|_2^2$
when the norm of $W_0$ increases.

\subsection{Nonlinear systems}

The previous analysis extends to the problem of finding 
the LQR law for a nonlinear system around an equilibrium
using data collected from the nonlinear system.
In fact, around an 
equilibrium a nonlinear system can be expressed via its first 
order approximation plus a reminder, which acts as 
a process disturbance for the linearized dynamics. \\[0.2cm]
Consider a smooth nonlinear system 
\begin{eqnarray} \label{nonl}
x(k+1) = f(x(k),u(k)) + \xi(k)
\end{eqnarray}
where $\xi$ is a process disturbance,
and let $(\overline x, \overline u)$ be a \emph{known} equilibrium pair, 
that is such that 
$\overline x = f(\overline x, \overline u)$. Thus, we can
rewrite the dynamics as
\begin{eqnarray} \label{nonl2}
\delta x(k+1) = A \delta x(k)  + B \delta u(k) + d(k)
\end{eqnarray}
with $\delta x := x- \overline x$, $\delta u := u-\overline u$, 
\begin{eqnarray*} 
A:=\left.\frac{\partial f}{\partial x}\right|_{(x,u)=(\overline x, \overline u)},\quad
B:=\left.\frac{\partial f}{\partial u}\right|_{(x,u)=(\overline x, \overline u)} \,.
\end{eqnarray*}
and with $d:=\xi+r$, where
$r$ accounts for higher-order terms and it has the property that is 
goes to zero faster than $\delta x$ and $\delta u$, namely we have
\begin{eqnarray} \label{eq:remainder}
r=R(\delta x, \delta u)\begin{bmatrix} \delta x \\ \delta u\end{bmatrix}
\end{eqnarray}
where $R(\delta x, \delta u)$ ia a matrix of smooth functions with the property that 
$R(\delta x, \delta u)$ goes to zero as $[ \delta x^\top \, \delta u^\top ]^\top$
goes to zero. Now,
if the pair $(A,B)$ defining the linearized 
system is stabilizable then a controller $K$ rendering $A+BK$ 
stable also exponentially stabilizes the equilibrium $(\overline x, \overline u)$ 
for the original nonlinear system. Thus, the analysis
in Theorem \ref{thm:main3} carries over directly to this case
(similar conclusions apply to Theorem \ref{thm:main4}).

\begin{corollary} \label{cor:nonlinear_stability} 
Consider a nonlinear system as in \eqref{nonl}, 
along with a known equilibrium pair $(\overline x, \overline u)$.
and let $K_{opt}$ be the optimal LQR controller of the 
system linearized around $(\overline x, \overline u)$.
Then, Theorem \ref{thm:main3} holds with \eqref{eq:system} replaced by \eqref{nonl2}. \qedp
\end{corollary}

\subsection{De-noising through averaging} \label{Subs:ave}

Several de-noising strategies can be adopted 
when the noise features are known, 
popular methods being the Singular Spectrum Analysis, the Cadzow algorithm,
and structured low-rank approximation \citep{Chu2003,Golyandina2001}.
Here, we discuss a simple de-noising strategy based on averaging
of \emph{ensembles} \citep{Wang1945}. \\[0.2cm]
Roughly, the idea is that
for signals affected by random noise the 
components due to noise
can be filtered out by taking an average of several signal ``cycles". 
This can be done by considering a single trajectory of length $T_*$
and cutting it into $N$ pieces of length $T$ (single trajectory ensemble)
or by taking $N$ measurements of length $T$
(multiple trajectory ensemble). We now  
elaborate on this idea considering the 
case of multiple trajectory ensembles. \\[0.2cm]
Given $N$ matrices $S^{(n)}$ with $n=1,\ldots,N$, let
\begin{eqnarray}
\overline S := \frac{1}{N} \sum_{n=1}^N S^{(n)}
\end{eqnarray} 
denote their average. For a given $N$, let
\begin{eqnarray}
x^{(n)}(k+1) = A x^{(n)}(k) + B u^{(n)}(k) + d^{(n)}(k)
\end{eqnarray} 
be the dynamics of \eqref{eq:system} over a generic 
experiment (cycle) $n$ with $n=1,\ldots,N$. Thus,
$x^{(n)}$, $u^{(n)}$ and $d^{(n)}$ are the state, input, and disturbance signals associated with the experiment $n$.
By linearity,
if we collect $T$ samples in each experiment the resulting tuples
\begin{eqnarray*}
\left( U_{0}^{(n)},D_{0}^{(n)},X_{0}^{(n)},X_{1}^{(n)} \right), \quad n=1,\ldots,N
\end{eqnarray*} 
satisfy the relation
\begin{eqnarray} \label{eq:recursive_ave}
\overline X_1 = A \overline X_0 + B \overline U_0 + \overline D_0 
\end{eqnarray} 
Hence, the average signals still provide a valid input-output 
system trajectory, meaning that all previous results
apply to this case without any modifications. \\[0.2cm]
For random noise, however, using \eqref{eq:recursive_ave}
can be advantageous with respect to using \eqref{eq:recursive}, that is 
one single experiment.
To see this, consider the case of $N$ (repeated)
experiments carried out with persistently exciting input signals
$u^{(n)} = u$ for all $n=1,\ldots,N$ and arbitrary initial states, 
and suppose that the noise realizations $d^{(n)}$ are
\emph{i.i.d.} with zero mean and covariance matrix $\sigma^2 I$.
Under these conditions \eqref{eq:recursive_ave} holds with 
$\overline U_0 = U_0$, \emph{i.e.}
$\overline X_1 = A \overline X_0 + B U_0 + \overline D_0$. This  
ensures that the average trajectory arises from a persistently exciting input, 
which is needed for having \eqref{rank.condition} fulfilled
(the average of persistently exciting signals need not result in 
a persistently exciting signal).
With this appraoch,
\eqref{eq:stab_practical} and \eqref{eq:stab_practicala} 
(thus \eqref{eq:constr_1} and \eqref{snr.V}) become 
easier to satisfy. In fact, 
\begin{eqnarray} \label{eq:approx_cov}
\frac{1}{T}\,D_{0}^{(n)} (D_{0}^{(n)})^\top \approx \sigma^2 I
\end{eqnarray} 
where the accuracy of the approximation increases with $T$
(the relation being exact in terms of \emph{expectation}).
Hence,
\begin{eqnarray} \label{eq:approx_cov_ave}
\overline D_0 {\overline D}_0^\top \approx
\frac{1}{N^2}\,\sum_ {n=1}^N D_{0}^{(n)} (D_{0}^{(n)})^\top
\approx \frac{T}{N}\,\sigma^2 I
\end{eqnarray} 
showing an approximate reduction by a factor of $N$ 
(indeed, this is nothing but a consequence of the fact that 
averaging $N$ \emph{i.i.d} realizations reduces the variance by a factor of $N$).
This procedure is illustrated in the numerical simulations 
which follow.

\begin{table*}[h!]
\tiny
\def\arraystretch{1.3}
\begin{tabular}{|r||c|c|c|c|c|c||c|c|c|c|}
\hline   
& \textbf{WGN} & \textbf{WGN}  & \textbf{WGN}  & 
\textbf{WGN}  & \textbf{WGN} & \textbf{WGN} & \textbf{Constant bias} & 
\textbf{Constant bias} & \textbf{Sine wave} & \textbf{Sine wave} \\
& $\sigma =  0.01$  & $\sigma =  0.03$ & $\sigma =  0.05$ & 
$\sigma =  0.1$ & $\sigma =  0.3$
& $\sigma =  0.5$ 
& $\overline \kappa =  0.05$ & $\overline \kappa =  0.1$ & 
$\overline \kappa =  0.05$ & $\overline \kappa =  0.1$ \\
& SNR\,\,$=$\,\,32.0  & SNR\,\,$=$\,\,22.5 & SNR\,\,$=$\,\,18.0 & 
SNR\,\,$=$\,\,12.0 & SNR\,\,$=$\,\,2.5
& SNR\,\,$=$\,\,-1.9 & SNR\,\,$=$\,\,23.3 & SNR\,\,$=$\,\,17.3 & SNR\,\,$=$\,\,28.7
& SNR\,\,$=$\,\,22.7 \\
\hline \hline
$\mathcal S $ for \eqref{lqr-forma} &  $\textbf{100\%}$  & ${97\%}$ & $95\%$ & $91\%$ 
& $83\%$ &  $\textbf{78\%}$ & $97\%$ & $\textbf{96\%}$ & $98\%$ & $96\%$ \\
$\mathcal S $ for \eqref{lqr-formb} &   $\textbf{100\%}$  & $\textbf{98\%}$ & $\textbf{96\%}$ 
&  $\textbf{93\%}$ & $\textbf{85\%}$ & $\textbf{78\%}$ &  $\textbf{98\%}$ & ${95\%}$ 
& $\textbf{100\%}$ & $\textbf{98\%}$ \\
\hline
$\mathcal M $ for \eqref{lqr-forma} &  $\textbf{0.0011}$ & $\textbf{0.0022}$ & 
$\textbf{0.0052}$ & $\textbf{0.0137}$ & $\textbf{0.0469}$ & ${0.0889}$ & $\textbf{0.0024}$ & 
$\textbf{0.0055}$ & $\textbf{0.0017}$ & $\textbf{0.0025}$ \\
$\mathcal M $ for \eqref{lqr-formb}  &  $0.1293$    & $0.0948$ &  $0.0757$ & $0.0433$ 
& $0.0498$ & $\textbf{0.0819}$ &  $0.1554$ & $0.2055$ & $0.1999$ & $0.2380$ \\
\hline 
$\mathcal V$ for \eqref{lqr-forma} &  $92\%$    & $75\%$ &  $50\%$ & $\textbf{11\%}$ 
& $0\%$ & $0\%$ & $36\%$ & $8\%$ & $38\%$ & $7\%$ \\
$\mathcal V$ for \eqref{lqr-formb} &  $\textbf{98\%}$    & $\textbf{81\%}$ &  $\textbf{51\%}$ 
& $6\%$ & $0\%$ & $0\%$ & $\textbf{67\%}$ & $\textbf{32\%}$ & $\textbf{71\%}$ & 
$\textbf{35\%}$ \\
\hline \hline
$\mathcal S_{ave}$ for \eqref{lqr-forma} &  $100\%$    & $100\%$ &  ${100\%}$ & ${100\%}$ 
& ${96\%}$  &  ${95\%}$ & $-$ & $-$ & $-$ & $-$ \\
$\mathcal M_{ave}$ for \eqref{lqr-forma} &  $0.0012$  & $0.0013$ &  $0.0013$ & $0.0014$ 
& $0.0034$  &  $0.0050$ & $-$ & $-$ & $-$ & $-$ \\
$\mathcal V_{ave}$ for \eqref{lqr-forma} &  $100\%$    & $99\%$ &  ${97\%}$ & ${94\%}$
& ${70\%}$  &  ${39\%}$ & $-$ & $-$ & $-$ & $-$ \\
 \hline
\end{tabular} 
\caption{Simulation results for $100$ random linear 
systems. For WGN, the last three rows report the simulation results
for \eqref{lqr-forma} in case of repeated experiments. 
Similar results are obtained for \eqref{lqr-formb}.
The values of SNR are in dB.} 
\label{sim:table}
\end{table*}

\section{Monte Carlo simulations} \label{Sec7}

In this section, we support our theoretical findings through 
simulations on linear and nonlinear systems. 

\subsection{Random linear systems} \label{Sec7}

We consider $100$ systems as in \eqref{eq:system} with $n=3$ and $m=1$,
under three types of noise: \emph{white Gaussian noise} (WGN),
constant bias and sinusoidal disturbances. In all the cases, we also consider
different levels of noise. For every type (and level) of noise
we test \eqref{lqr-forma} and \eqref{lqr-formb} in all the
systems. 
Numerical simulations have been carried out in Matlab. 
For each experiment, 
we choose the entries of the matrices $A$ and $B$ and of
the initial state from a normal distribution with zero mean
and unit variance, abbreviated by $\mathcal N(0,1)$ 
(command \texttt{randn}). For each experiment,
the controller was designed using $T=20$ samples
generated by applying an input signal $u \sim \mathcal N(0,1)$ 
(by Lemma \ref{lem:willems} condition \eqref{rank.condition} requires 
a minimum of $7$ samples). \\[0.2cm]
WGN has been generated 
taking $d \sim \mathcal N(0,\sigma^2 I)$, where $\sigma$ represents
the standard deviation. We varied $\sigma$ considering different 
scenarios of the signal-to-noise (SNR), computed (command \texttt{snr}) 
by comparing the variables $Bu$ (signal) and $d$ (noise).
This SNR measures how much noise enters the system relatively 
to the intended input signal.
Constant bias was chosen by applying to each input channel 
a value $\kappa$ taken from a uniform distribution 
in $(-\overline \kappa,\overline \kappa)$. Finally, sinusoidal disturbance was chosen 
by applying to each input channel a signal 
$\kappa \sin(k)$ with $\kappa$ given as above. \\[0.2cm]
We denote by $\mathcal S$
the percentage of times we get a stabilizing controller.
We also compute the performance gap between the controller
found via \eqref{lqr-forma} and \eqref{lqr-formb} and the optimal one.
Specifically, for each type (and level) of noise, we
let $\overline K{}^{(k)}$ and $K_{opt}^{(k)}$ with $k =1,\ldots,100$
denote the controller found via \eqref{lqr-forma} or \eqref{lqr-formb} and the optimal one 
for the $k$-th experiment, and let 
\begin{eqnarray*}
\mathcal E_k := \frac{\| \mathscr T(\overline K{}^{(k)}) \|_2^2 
- \| \mathscr T(K_{opt}^{(k)}) \|_2^2}
{\| \mathscr T(K_{opt}^{(k)}) \|_2^2} 
\end{eqnarray*}
represent the relative performance error. 
We denote by $\mathcal M$ the median of $\mathcal E_k$ through all the experiments
that return a stabilizing controller. Each type (and level) of 
noise was tested with the same set 
of plant matrices and inputs. Finally, we denote
by $\mathcal V$ the percentage of times we 
infer stability via \eqref{eq:stab_practical}
and \eqref{eq:stab_practicala} assuming some prior knowledge on $d$.
As for WGN, we selected $\delta$ in \eqref{eq:stab_practical}
and \eqref{eq:stab_practicala}
by taking $\hat \sigma = 1.5 \sigma$ ($50\%$ overestimate of $\sigma$) and by setting
$\delta=\sqrt{T} \hat \sigma$ (\emph{cf.} \eqref{eq:approx_cov}). 
As for constant and sinusoidal disturbances, we consider a worst-case estimate
$\hat D_0 = \overline \kappa \mathbf 1_{n \times T}$ where $\mathbf 1_{n \times T}$
is the ${n \times T}$ matrix of all ones, yielding
$\delta = \sqrt{T n} \overline \kappa$. These values of $\delta$
give a correct over-approximation of the norm of $D_0$
in all the experiments. These values of $\delta$ are also used to 
implement \eqref{lqr-formb}. Specifically, we implemented \eqref{lqr-formb}
by first computing the smallest $\mu^2$ such that
$\delta^2 I \preceq \mu^2 RR^\top$ with the choice $R=X_1$ 
and then by performing a line search on $\eta_1$.
The choice $R=X_1$ has robust stability interpretations
\citep[Section V]{dept2020} and proved effective in the simulations. \\[0.2cm]
For solving \eqref{lqr-forma} and \eqref{lqr-formb} we used CVX \citep{cvx}.
The results are reported in Table \ref{sim:table}, and they can be summarized
as follows: 
\begin{enumerate}[1.]
\item In all experiments, both methods \eqref{lqr-forma} and \eqref{lqr-formb} 
perform well for reasonable values of the SNR ($\geq25$dB) as well as for 
low-medium SNR values in the range $(10,20)$dB. 
The method \eqref{lqr-forma} performs better 
in terms of relative error but is slightly less robust, in line with the discussion 
in Section \ref{subsec:robustified}.
For very low SNR ($\leq5$dB) the performance of both methods drop. 
We note (not reported in Table \ref{sim:table}) that both
methods settle to $\mathcal S=76\%$ for SNR $\leq-5$dB regardless of $\sigma$. 
This happens since $76\%$ of systems are open-loop stable, and 
$K=0$ is feasible for both methods when $U_0$, $X_0$ and
$X_1$ have full-row rank. In this case, both methods select $Q$ such that
$U_0Q=0$, $X_1Q=0$ and $X_0Q=I$, leading to $P=I$ and $K=0$.
\item For both \eqref{lqr-forma} and \eqref{lqr-formb}, robustness to 
noise can be further enhanced 
by adding a weight $\alpha >1$ to the term ${\rm trace}(V)$, so as to
favour robustness over accuracy relative to $K_{opt}$ 
(\emph{cf.} Section \ref{Sec5}). For instance, for WGN with $\sigma=0.1$ the 
program \eqref{lqr-forma}
achieves $\mathcal S=96\%$ with $\alpha=10$,
but at the expense of a reduced performance $\mathcal M=0.0380$.
Using a weight $\alpha >1$ can be beneficial also for
stability inference (quantity $\mathcal V$) since smaller $V$
render \eqref{eq:stab_practical} and \eqref{eq:stab_practicala} easier to fulfil. 
For instance, under the same conditions as above $\mathcal V$ increases 
from $11\%$ to $46\%$.
\item For WGN, robustness can be increased also by
averaging trajectories from multiple experiments (\emph{cf.} Section \ref{Subs:ave})
This is advantageous with respect to adding a penalty on ${\rm trace}(V)$ because
no performance losses are introduced. To emphasize this point,
the last three rows of Table \ref{sim:table}
report the results with $N=100$ repeated experiments 
for each system, although $N=10$
suffices to get $\mathcal S=96\%$ with median 
relative error $\mathcal M=0.0034$ for $\sigma=0.1$, and
$\mathcal S=90\%$ with $\mathcal M=0.0296$ for $\sigma=0.5$.
\item With stable dynamics increasing $T$ is usually beneficial for performance. 
From a theoretical viewpoint, this is due the fact that increasing $T$
reduces the term ${\rm trace}(V_0)$, thus the relative error
(\emph{cf.} Section \ref{Subsec_error}). 
With unstable dynamics this advantage is offset by the fact that the noise effect 
amplifies, and this renders stability more difficult to achieve. In fact, we observed 
that decreasing $T$ actually gives an increase of $\mathcal S$ in almost all scenarios
since in this case stabilization of the unstable systems becomes easier.
(for instance, with $T=10$ we obtain $\mathcal S=82\%$ for WGN with $\sigma=0.5$).
\end{enumerate}

We have also tested our methods on the Laplacian system considered in
\citep{Dean2018journ}. With \eqref{lqr-forma}, under the same setting (input and noise
in $\mathcal N(0,1))$, an average of $N=10$ trajectories of 
length $T=20$ is sufficient to get $\mathcal S=100\%$ with $\mathcal M=0.6569$
over $100$ experiments made by randomly changing input and noise patterns.
To further decrease $\mathcal M$ one needs to increase $T$ (and $N$).
In this case, increasing $T$ does not bring issues since the dynamics
are mildly unstable and the input signals have zero mean.

\subsection{Nonlinear inverted pendulum}

Consider the Euler discretization of an inverted pendulum. The
system is as in \eqref{nonl} with 
\begin{eqnarray*}
f(x,u)= \left[
\begin{array}{l}
x_1 + \Delta x_2 \\[0.1cm]
\displaystyle 
\frac{\Delta  g}{\ell} \sin x_1 + \left( 1 - \frac{\Delta  \mu}{m \ell^2} \right) x_2
+\frac{\Delta }{m \ell^2}u\\
\end{array}
\right]
\end{eqnarray*}
where $\Delta$ is the sampling time, 
$m$ is the mass, $\ell$ is the distance from the base to the center of mass of the
balanced body, $\mu$ is the coefficient of rotational friction, and $g$ is the acceleration due to gravity. 
The states $x_1, x_2$ are the angular position and velocity, respectively, $u$ is the applied torque.
The system has an unstable equilibrium in 
$(\overline x, \overline u)=(0,0)$ corresponding to the pendulum upright position
so that $\delta x= x$ and $\delta u=u$.
We assume that the parameters are 
$\Delta= 0.01$, $m=\ell=1$, $\mu=0.01$, and $g=9.8$. \\[0.2cm]
We made $100$ experiments by considering initial conditions 
in $\mathcal N(0,0.1)$, corresponding to an initial displacement from
the equilibrium of about $\pm10^\circ$, and $u \sim \mathcal N(0,1)$. The results are in line with
the previous ones. In particular, when $\xi=0$ (the only disturbance source
is the nonlinearity) we obtain $\mathcal S=100\%$ with $\mathcal M=0.0356$
using \eqref{lqr-forma} with
trajectories of length $T=20$. We also considered 
the case of WGN noise affecting the velocity dynamics, \emph{i.e.} 
with $u$ replaced by $u+\xi$ with $\xi \sim \mathcal N(0,\sigma)$. In this case,
we obtain $\mathcal S=100\%$ for $\sigma \leq 0.1$ (SNR $\geq20$dB)
up to $\mathcal S=12\%$ for $\sigma = 1$ (SNR $\approx 0$dB). 
Similar results are obtained with \eqref{lqr-formb} and under different
settings, that is with different types of noise and samples $T$. Since the equilibrium 
is unstable, reducing $T$ can be beneficial for values of $u$ and $\xi$ that 
steer the system far from the equilibrium (for instance, using $T=10$ we get 
$\mathcal S=36\%$ for WGN $\sigma = 1$). As for linear systems, at the expense of 
reduced performance, robustness can be enhanced by adding a 
weight $\alpha >1$ to the term ${\rm trace}(V)$ 
(for instance, setting $\alpha = 10$ we obtain $\mathcal S=64\%$ for WGN with $\sigma = 1$).

\section{Concluding remarks} \label{Sec8}

The design of (optimal) controllers from noisy data is a very challenging 
and largely unsolved problem. In this paper we took some steps in
this direction for the LQR problem. By resorting to a convex SDP formulation 
of the LQR problem, we proposed two novel methods
that explicitly account for noise through an augmented cost function which favours 
noise-robust solutions.
Both method provides finite sample stability guarantees,
and do not require specific noise models 
such as the noise being white.\\[0.2cm]
A great leap forward would come from extending 
the ideas of this paper to incorporate state and input \emph{safety} constraints
\citep{Zeilinger2019}. At the moment of writing, we aim at tackling this challenge
using concepts and tools from \emph{set-invariance} control. 
For stabilization problems with no optimality requirements, recent results 
have shown that data-based formulations of set-invariance properties
can be efficiently cast as linear programs, and they can handle 
noisy data \citep{Bisoffi2019}.

\appendix

\section{Appendix}

\emph{Proof of Lemma \ref{lem:tech1}}. The proof 
follows the same logical steps as 
\citep[Proposition 3.13]{scherer-LMI-book} given
for the model-based approach. Here, we consider a data-based version.
Since $X_{0} Q = P$ and $K=U_0QP^{-1}$ we have
\begin{eqnarray*}
\left[ \begin{array}{c} K \\ I \end{array} \right] =
\left[ \begin{array}{c} U_0 \\ X_0 \end{array} \right] QP^{-1} = W_0QP^{-1}
\end{eqnarray*}
This implies $A+BK=(X_1-D_0) QP^{-1}$.
%\begin{eqnarray*}
%A+BK = \left[ \begin{array}{cc} B & A \end{array} \right] 
%W_0 QP^{-1} = (X_1-D_0) QP^{-1} 
%\end{eqnarray*}
Hence, the first constraint in \eqref{lqr-form_ideal}
is equivalent to $S\preceq 0$ where
\begin{eqnarray} \label{eq:S}
S := (A+BK) P (A+BK)^\top  - P + I 
\end{eqnarray}
Hence $K$ is stabilising. As for the second part of the 
claim, since $S$ is symmetric there exists a matrix $\Xi$ such that
$S + \Xi \Xi^\top = 0$. 
Thus, $P$ coincides with the controllability Gramian
of the extended system 
\begin{eqnarray} \label{eq:closed_ext}
\left[
\begin{array}{c}
x(k+1) \\ z(k)
\end{array}
\right]  = 
\left[
\begin{array}{c|c}
A+BK & 
\left[ \begin{array}{ccc} I & & \Xi \end{array} \right]  \\ \hline 
\left[ \begin{array}{c} I \\ K \end{array} \right]  & 0
\end{array}
\right] \left[
\begin{array}{c}
x(k) \\ d(k) \\ \xi(k)
\end{array}
\right] 
\end{eqnarray}
with $\xi$ and additional input.  
Let us call $\mathscr T_e(K)$ the transfer function of \eqref{eq:closed_ext}.
By definition of $\mathcal H_2$-norm, we have
\begin{eqnarray*}
\|\mathscr T_e(K)\|_2^2 \,\, % &=&\,\, {\rm trace} \left( (I + K^\top K) P \right) \\
&=&\,\, {\rm trace} \left( P \right) + {\rm trace} \left( K P K^\top \right) \\
&=&\,\, {\rm trace} \left( P \right) + {\rm trace} \left( U_0 Q P^{-1} Q^\top U_0^\top \right) \\
&\leq&\,\, {\rm trace} \left( P \right) + {\rm trace} \left( L \right) 
\end{eqnarray*}
We conclude that $\|\mathscr T(K)\|_2^2 \leq 
{\rm trace}\left( P \right) + {\rm trace} \left( L \right)$ since
$\mathscr T_e(K) = [ \mathscr T(K) \,\,\,\, \mathscr T_\Xi(K) ]$,
where $\mathscr T_\Xi(K)$ is the transfer function of the system
\begin{eqnarray} \label{eq:closed_residue}
\left[
\begin{array}{c}
x(k+1) \\ z(k)
\end{array}
\right]  = 
\left[
\begin{array}{c|c}
A+BK & \Xi \\ \hline 
\left[ \begin{array}{c} I \\ K \end{array} \right]  & 0
\end{array}
\right] \left[
\begin{array}{c}
x(k) \\ \xi(k)
\end{array}
\right] 
\end{eqnarray}
This concludes the proof. \qedp

\emph{Proof of Lemma \ref{lem:tech2}}. 
Consider any stabilising controller and 
denote by $P$ the controllability Gramian
associated with the closed-loop system \eqref{eq:closed},
which solves $S=0$ with $S$ as in \eqref{eq:S}.
Consider the system of equations \eqref{eq:constG}
in the unknown $G$, and which admits a solution under \eqref{rank.condition}. 
In particular, pick 
\begin{eqnarray*}
G_* := W_0^\dag 
\left[ \begin{array}{c}
 K \\ I
\end{array} \right]
\end{eqnarray*}
where $\dag$ denotes the right inverse.
Starting from the matrices $P$ and $G_*$, define
$Q := G_* P$, $L := U_0 Q P^{-1}Q^\top U_0^\top$ and
$\gamma := {\rm trace} (P) + {\rm trace} (L)$.
Hence, $(\gamma,Q,P,L)$ is feasible for \eqref{lqr-form_ideal}.
Moreover, $K=U_0 Q P^{-1}$.
Finally, the last part of the claim follows from
\begin{eqnarray*}
\|\mathscr T(K)\|_2^2 \,\, % &=&\,\, {\rm trace} \left( (I + K^\top K) P \right) \\
&=&\,\, {\rm trace} \left( P \right) + {\rm trace} \left( K P K^\top \right) \\
% &=&\,\, {\rm trace} \left( P \right) + {\rm trace} \left( U_0 Q P^{-1} Q^\top U_0^\top \right) \\
&=&\,\, {\rm trace} \left( P \right) + {\rm trace} \left( L \right) 
\end{eqnarray*}
This gives the claim. 
As it emerges from the proof, to any stabilizing controller we can associate an
infinite number of tuples $(\gamma,Q,P,L)$ feasible for \eqref{lqr-form_ideal} 
with $Q=G_* P+Q_\sim$,
where $Q_\sim$ is any matrix in the right kernel of $W_0$.
\qedp

\bibliographystyle{plainnat} 

\bibliography{manuscript}

\end{document}